\documentstyle[prl,epsf,twocolumn,aps]{revtex}

\begin{document}

\title{Scale Dependence of Intermittency Exponents 
in Developed Hydrodynamic Turbulence}

\author{Vladimir~M.~Malkin}
\address{Institute for Advanced Study, Princeton, NJ 08540 \&
Budker Institute of Nuclear Physics, Russia
}

\date{\today}

\maketitle

\begin{abstract}
The scale dependent intermittency exponents
in developed hydrodynamic turbulence are calculated assuming
a natural hierarchy of correlations in the turbulence.
The major correlations are taken into account explicitly,
while the remaining small correlations are considered as 
perturbations. The results agree very well with the currently
available experimental data.
\end{abstract}

PACS-number: 47.27

\bigskip
$\bullet$ {\bf INTRODUCTION:} \indent
The energy dissipation in developed hydrodynamic turbulence
is strongly non-uniform in space (and time). 
This property is usually referred to as an intermittency.
A reason for intermittency was already present in the
original Kolmogorov-Obukhov model of turbulence, as it was uncovered
by the well-known comment of Landau \cite{kl}. 
The major process in developed
turbulence is the  energy transfer 
from turbulent motions of large spatial scales
(where the energy is basically contained) to the smallest scales 
(where the dissipation occurs). The transferred energy flux becomes
more and more spatially non-uniform as it goes to smaller
and smaller scales, which ultimately leads to pronounced intermittency.
A reason for this is that the velocity field at smaller scales is
capable of changing in time much faster than at larger scales.
It implies that a quasi-stationary (in statistical sense)
picture of turbulence has time enough to be established at smaller
scales before a noticeable change in the local large-scale flow
occurs. Therefore, the statistical characteristics of smaller-scale motions
depend just on the local (in space and time) larger-scale motion.
Then, there is no smoothing mechanism
for spatial fluctuations in the density
of energy flux and those are
amplified as the flux goes from larger to smaller scales.
The quantitative description for such an
intermittent amplification 
has not been developed yet because of the complexity
of the problem. 

An idea is suggested below on how to 
overcome the difficulties, assuming that not all of
the correlations causing the intermittency are  equally
important. It appears to be possible to derive and solve
explicitly an equation that takes into account the major
correlations. After
a renormalization, a local flux-conservation constrain 
leads to an equation that has (in addition the Kolmogorov
solution) an intermittent solution. The latter agree
very well with experiment, which supports the employed
assumptions about hierarchy of correlations.

\bigskip
$\bullet$ {\bf ENERGY FLUX AND DISSIPATION:} \indent
The problem of intermittency can be considered quantitatively
in the terms of energy and flux densities in $(\vec k , \vec r )$-space. 
Provided the energy is pumped  at the smallest
wave numbers $k\sim k_0$, damped at the largest 
$k\sim k_d\gg k_0$ and
there is neither pumping nor damping at all intermediate $k$,
the densities of energy and flux satisfy in the latter --
inertial -- range the continuity equation. It can be averaged
over directions of $\vec k$ and over some domain of fixed shape 
in $\vec r$-space. The result contains a surface integral
(over the boundary of domain), that
describes the spatial transfer of energy from the domain to its neighborhood,
and an integral over the domain, that represents the averaged 
density of energy flux from smaller to larger $k$. The latter quantity is
denoted below as $\epsilon(\vec R,x,k)$, where $\vec R$ and $x$
define the spatial position  and linear size of the domain;
the time variable and variables describing the
shape of domain are omitted for simplicity.  
One could expect that the average flux density $\epsilon(\vec R,x,k)$
tends at $k\gg 1/x$ (but $k$ is inside the inertial range)
to a $k$-independent limit. Such a limit, if it exists,
should be identified with the average density of
energy dissipation in given domain
at given moment of time, which quantity
is denoted below as $\epsilon(\vec R,x)$.

Statistical properties of random fields $\epsilon(\vec R,x,k)$
and $\epsilon(\vec R,x)$ may be sensitive to the shape of domain used for 
the averaging of energy flux and dissipation densities. 
Therefore, for comparison with experiments, the
same shapes as there must be employed in the theory as well. 
Because of the technical reasons, the experiments usually deal with
the so called ``one-dimensional cuts'', obtained by means
of small probes that record the longitudinal velocity of fluid 
in fixed points passed by the stream.  
In some recent experiments the transverse 
component of velocity is measured, but the currently
available database contains primarily records of
the longitudinal velocity. 
When such data are processed, the density of energy dissipation
rate is substituted usually by the square of the velocity
derivative along the record.

\bigskip
$\bullet$ {\bf BASIC EQUATIONS:} \indent
In order to describe the amplification of fluctuations
in density of energy flux from larger to smaller scales,
consider the following conditional probabilities.

Let $G_n(M)dM$ be the probability to find the flux
$\epsilon(\vec R,x/2^n,2^n /x)$ in the range
$ (M,M+dM)\epsilon$, provided $\epsilon(\vec R,x,1/x)=\epsilon$.
The ensemble averaging means here and further the $\vec R$-space averaging.
The probability density $G_n(M)$ does not depend on $x$ 
(inside the inertial range). It is a scale-invariance hypothesis.
The $\epsilon$-dependence of  $G_n(M)$ (which would imply a correlation 
between the local amplification coefficient $M$ and energy flux
density $\epsilon$) is assumed to be negligible in a zero-order approximation.
When correlation between the consequent
amplification coefficients is also negligible, the moments
$G_{n,q}\equiv\int dM M^q G_n(M)$ depend on $n$
exponentially:
\begin{equation}
G_{n,q}=G_{1,q}^n\equiv G_q^n~.
\label{1}
\end{equation}
Let $g_n(M)dM$ be the probability to find the flux
$\epsilon(\vec R,x,2^n /x)$ in the range
$(M,M+dM)\epsilon $, provided $\epsilon(\vec R,x,1/x)=\epsilon$.
The consequent amplification coefficients of this kind are
strongly correlated, so that $n$-dependence of $g_n(M)$ is not trivial.
There is, however, a relation between the probability densities $g_n(M)$  
and $G_1(M)\equiv G(M)$. To find it, consider the ``one dimensional
cut'' of length $x$ as the sum of its halves (which length is 
$x/2$). The conditional probability to find the flux
$\epsilon(\vec R,x/2,2/x)$ in the range
$(M,M+dM)\epsilon $, provided $\epsilon(\vec R,x,1/x)=\epsilon$,
is $G(M)dM$. It relates to each half of the segment $x$, when
the dependence on the position of the half inside
the segment is negligible. The amplification coefficients for the halves 
virtually do not affect each other, because of the strongly
fluctuating short-scale spatial transfer of the energy in
directions transverse to the segment. Then, the following 
recurrent relation holds up to small corrections:
\begin{eqnarray}
&g_n(M)=\int dM_1 dM_2 dM_3 dM_4 G(M_1) G(M_2)\times \nonumber \\
&g_{n-1}(M_3) g_{n-1}(M_4)\delta \left( M - (M_1 M_3+ M_2 M_4)/2 \right) ~.
\label{2}
\end{eqnarray}
It can be reduced to the recurrent relation for moments
$g_{n,q}\equiv\int dM M^q g_n(M)$, that looks as
\begin{equation}
g_{n,q}=\sum_{q_1=0}^{q_1=q}\frac{ q! G_{q_1} G_{q-q_1} g_{n-1,q_1} 
g_{n-1,q-q_1}}  {(q_1)!(q-q_1)!2^q}~.
\label{3}
\end{equation}
Eq.(\ref{3}) with ``initial conditions''  $g_{0,q}=g_{n,0}= 1$  
allows one to express $g_{n,q}$ in the terms of 
$G_q$ for all positive integer $n$ and $q$. 

The $n$-dependent moments of energy flux densities
averaged over segment $x$ and over its part $x/2^n$ 
at the same wave number $k_n=2^n /x$ would be 
equal to each other for a spatially uniform flux. 
For an intermittent energy flux, the ratio  
\begin{equation}
G_{n,q}/g_{n,q}\equiv 2^{n\mu_{n,q}}
\label{4}
\end{equation}
increases in $n$ (which indicates the amplification of fluctuations).  
According to (\ref{4}) and (\ref{1}), the ``intermittency exponents'' 
$\mu_{n,q}$ are given by
\begin{equation}
\mu_{n,q}=\log_2{G_q} -n^{-1}\log_2{g_{n,q}}~.
\label{5}
\end{equation}

\bigskip
$\bullet$ {\bf EXPLICIT SOLUTION OF EQ.(\ref{3}):}\hskip10pt
Equation (\ref{3}) can be solved explicitly, and thus
the ``intermittency exponents'' $\mu_{n,q}$ can be expressed
in the terms of $n$-independent functions $G_q$.
Provided the latter satisfy restriction $G_q < 2^{q-1}$,
the solution $g_{n,q}$ tends to finite limit $g_q$ as
$n$ tends to infinity. Then, the ``intermittency exponents'' 
$\mu_{n,q}$ tend to $\mu_q=\log_2{G_q}$ as $n$ tends to infinity,
and the above restriction on $G_q$ is equivalent to $\mu_q < q-1$.
Explicit formulas for $g_{n,q}$ that solve (\ref{3})
can be derived consequently for  $q=1,2,...$.

For $q=1$, taking into account that $G_0=G_1=1$, one gets from
(\ref{3}) $g_{n,1}=1$, and hence $\mu_{n,1}=0$.

For $q=2$ the solution of (\ref{3}) is
\begin{eqnarray}
&g_{n,2}=g_2+c_{2}\,\left(G_2/2\right)^n\, ;\nonumber\\
&g_2=1/\left(2-G_2\right)\,, \;\;  c_{2}=1-g_2\,.
\label{6}
\end{eqnarray}
For $q=3$ the solution is
\begin{eqnarray}
&g_{n,3}=g_3+c_{3}\,\left(G_3/4\right)^n+c_{32}\,\left(G_2/2\right)^n\, ;\nonumber\\
&g_3=\frac{3G_2g_2}{4-G_3}\,,\;\;c_{32}=\frac{3c_2G_2}{2G_2-G_3}\,, \;\;  
c_{3}=1-g_3-c_{32}\,.
\label{7}
\end{eqnarray}
For $q=4$ it is
\begin{eqnarray}
&g_{n,4}=g_4+c_{4}\,\left(G_4/8\right)^n+c_{43}\,\left(G_3/4\right)^n+
\nonumber\\
&+c_{42}\,\left(G_2/2\right)^n+c_{422}\,\left(G_2/2\right)^{2n}\, ;\nonumber\\
&g_4=\frac{4G_3g_3+3G_2^2g_2^2}{8-G_4}\,,\;\;
c_{43}=\frac{4c_3G_3}{2G_3-G_4}\,,\nonumber\\   
&c_{42}=\frac{4c_{32}G_3+6c_2G_2^2g_2}{4G_2-G_4}\,, \;\;
c_{422}=\frac{3c_2^2G_2^2}{2G_2^2-G_4}\,, \nonumber\\
&c_{4}=1-g_4-c_{43}-c_{42}-c_{422}\,.
\label{8}
\end{eqnarray} 
For $q=5$:
\begin{eqnarray}
&g_{n,5}=g_5+c_{5}\left(G_5/16\right)^n+c_{54}\left(G_4/8\right)^n
+c_{53}\left(G_3/4\right)^n
\nonumber\\
&+c_{52}\,\left(G_2/2\right)^n+c_{532}\,\left(G_2G_3/8\right)^{n}
+c_{522}\,\left(G_2/2\right)^{2n}\, ;\nonumber\\
&g_5=5\frac{G_4g_4+2G_2G_3g_2g_3}{16-G_5}\,,\nonumber\\ 
&c_{54}=\frac{5c_4G_4}{2G_4-G_5}\,,\;\;  
c_{53}=5\frac{c_{43}G_4+2c_3G_3G_2g_2}{4G_3-G_5}\,, \nonumber\\
&c_{52}=5\frac{c_{42}G_4+2c_2G_3G_2g_3+2c_{32}G_3G_2g_2}{8G_2-G_5}\,,\nonumber\\
&c_{532}=\frac{10c_2c_3G_2G_3}{2G_2G_3-G_5}\,, \;\;
c_{522}=5\frac{c_{422}G_4+2c_2c_{32}G_3G_2}{4G_2^2-G_5}\,,\nonumber\\
&c_{5}=1-g_5-c_{54}-c_{53}-c_{52}-c_{532}-c_{522}\,.
\label{9}
\end{eqnarray} 
 
The similar (but longer) formulas exist for $q=6,7,8,...\,$.
In particular, the $n$-independent function $g_q$ is given in general by
\begin{equation}
g_q=\sum_{q_1=1}^{q_1=q-1}\frac{q!G_{q_1} G_{q-q_1} g_{q_1} g_{q-q_1}}
{(q_1)!(q-q_1)!(2^q-2G_q)}  ~.
\label{11}
\end{equation}

Consider now the alternative option $G_q>2^{q-1}$.
Under such a condition, $g_{n,q}$ does not tend to a finite limit as
$n$ tends to infinity, but increases exponentially like
$g_{n,q}\propto (G_q/2^{q-1})^n$. 
It follows then from (\ref{5}) that the intermittency exponent 
$\mu_{n,q}$ tends to
$q-1$ as $n$ tends to infinity  (and one can see that
$\mu_{n,q}$ approaches to $q-1$ from below). There is a simple
physical interpretation of such a behavior. It indicates
that the strongest fluctuations
of $\epsilon (\vec R, x)$ increase faster than $1/x$
as $x$ tends to zero (so
that the density of energy dissipation rate has 
singularities non-integrable over the one-dimensional
cuts).  According to the estimate
$\epsilon (\vec R, x) \sim {\tilde v(\vec R, x)}^3/x $
(where $\tilde v(\vec R, x)$ is a smoothed velocity variation  in
a fluctuation of size $x$ around point $\vec R$), 
it implies that the sweep $\tilde v (x)$ of 
$\tilde v (\vec R,x)$ variation in $\vec R$-space   
increases when the fluctuation  size $x$ tends to zero.
Thus, the alternative $G_q>2^{q-1}$ 
corresponds to blow-up of velocity itself,
rather than just its gradients, in Euler fluid. 

In spite of many efforts applied to the problem,
there is no clear theoretical answer yet to the question,
whether velocity blows-up or not in Euler fluid. 
Practical answer can be extracted from the experimental
data on intermittency, provided the collected statistics 
is sufficient to detect the strongest blow-up. The currently 
available database supports assumption $\mu_q < q-1$ for
$q=2,3,4,...\,$. 

\bigskip
$\bullet$ {\bf DEGENERATE SOLUTIONS:}\indent
The above formulas for $g_{n,q}$ contain all possible items
of the kind $(G_q/2^{q-1})^n\, ,\; q=2,3,... $ and their products.
There is however a special class of solutions that contain
only the items with $q=2$. It corresponds to the physical situation when
only one major kind of correlations exists, that dominates over all
other correlations. For such degenerate solutions, all quantities
$G_q$ (or $\mu_q=\log_2{G_q}$) can be expressed in the terms
of $\mu_2$. In the limit of infinitely small $\mu_2$, the 
expression looks as
\begin{equation}
\mu_q\approx\mu_2\,q(q-1)/2~.
\label{12}
\end{equation}
The dimensionality $D_q\equiv 3+\mu_q-q\mu'_q$ corresponding to (\ref{12}) is
$D_q\approx 3-\mu_2\,q^2/2~$.
It would turn into zero at $q=q_M\approx (6/\mu_2)^{1/2}$, if (\ref{12}) would
be applicable there, -- which is not so, as the applicability condition
$\mu_q\ll 1$ is violated at $q\sim q_M$. In fact, the applicability
condition is a little bit more soft quantitatively, it is $G_q-1\ll 1$.
The slightly improved formula (\ref{12}) is
\begin{equation}
G_q-1\approx (G_2^{q-1}-1)\,q/2~.
\label{14}
\end{equation}
It appears to be of a reasonable accuracy 
for real turbulence up to $q=5$ (see more accurate formulas below 
for the comparison).

\bigskip
$\bullet$ {\bf INTERMITTENCY EXPONENTS:}\indent 
For a moderately small $\mu_2$ that occurs in real turbulence,
the above formulas require a modification. An appropriate
formula for $\mu_q$ is derived from very general heuristic
reasons in \cite{vm} (where the major mechanism of intermittency
is supposed to be blow-up of velocity gradients in 
Euler fluid, leading to point singularities). 
This formula looks as
\begin{eqnarray}
&\mu_q=\gamma\, q \left[{\rm arsh}\left(q/a\right)-
{\rm arsh} \left(1/a\right)\right]~,\quad q\leq q_M~,
\label{15}\\
&\mu_q=\alpha_M\,q-3\, ,\quad (\alpha_M=\mu'_{q_M})\, ,\quad q\geq q_M~;
\label{16}\\
&q_M={3/\gamma}\left[ 1/2 +\left(1/4+
  \left(a\gamma/3\right)^2 \right)^{1/2} \right]^{1/2}~,
\label{17}
\end{eqnarray}
where ${\rm arsh}\xi=\ln(\xi+\sqrt{\xi^2+1})$ is the hyperbolic arcsin,
prime signifies $q$-derivative, $\gamma$ and $a$ are some numbers.
For $a\gg q$, (\ref{15}) reduces to (\ref{12}) with $\mu_2=2\gamma/a$.

There is a connection between parameters $\gamma$ and $a$, so that
$\mu_q$ can be expressed in the terms of $\mu_2$ for realistic values
of the latter parameter as well. The extra condition
has the form
\begin{equation}
F(\alpha_M,\alpha_0)\equiv 2^{3-\alpha_0}- 2^{\alpha_M}-7=o(1)~
\label{18}
\end{equation}
(which is easy to understand in the framework of a cubic shell model).
The quantity $\alpha_q=\mu'_q$ can be referred to as
the ``singularity exponent''. 
When $q$ increases from zero to $q_M$,
$\alpha_q$ increases from its smallest value $\alpha_0 <0$
to its largest value $\alpha_M=\alpha_{q_M}$. 
The exponent $\alpha_M$ corresponds to the strongest possible
amplification of $\epsilon$ in a ``hot spot'',
i.e., to the strongest blow-up in Euler fluid. 
The exponent $\alpha_0 <0$ corresponds to the background that goes down
to compensate for amplification of $\epsilon$ in the hot spot
and to secure the energy flux conservation. The flux conservation
condition around the strongest hot spot gives 
\begin{equation}
F(\alpha_M,\alpha_0)\approx\mu_2^2~.
\label{19}
\end{equation}

For comparison between the theory and experiment,
the above definition of intermittency exponents $\mu_{n,q}$
(selected to get explicit analytical formulas) must be slightly
modified and also small effects neglected in the
derivation of eq.(\ref{3}) must be taken into account.
The different definitions of intermittency exponents can
be identified with each other, as long as only scale-independent
intermittency exponents $\mu_q$ (corresponding to
infinite scale ratio) are considered. The really measured intermittency
exponents should be considered taking into account the finite
scale-ratio effects which are sensitive to the definition.
In particular, the above definition must be distinguished from that
used in \cite{ppn}. The latter deals with
the amplification coefficients $M(\vec R,x,y)\equiv 
\epsilon(\vec R,y)/\epsilon(\vec R,x)$, and defines
intermittency exponents in a way equivalent to
\begin{equation}
\overline{M(\vec R,x,x/2^n)^q}=2^{n\mu_{n,q}}~.
\label{20}
\end{equation}
Here bar signifies the $\vec R$-space averaging.
The result does not depend on $x$ due to the scale invariance.

The modification can be neglected for infinitely small $\mu_2$.
For a moderately small $\mu_2$, the quantities $\mu_{n,q}$ defined
by (\ref{20}) are expressed in the terms of $\mu_2$ by
\begin{equation}
\mu_{n,q}\approx\mu_q -{\sqrt{C}\over n}\left(\log_2{g_{n,q}}\right)^C,\;\; 
C\approx 1+\ln(1-\mu_2)
\label{21}
\end{equation}
(where $\mu_q$ is given by (\ref{15})-(\ref{19}) and $g_{n,q}$
is the above exact solution of eq.(\ref{3}) with $G_q=2^{\mu_q}$).
Formula (\ref{21}) is valid for $n=1,2,...$ and $q=2,3,...$.

Noteworthy, that small correlations between the local values of energy flux
density $\epsilon$ and amplification coefficient $M$ 
lead to a non-zero intermittency exponent
$\mu_{n,1}$ (which otherwise is zero and which usually is
assumed to be zero). It follows from the  definition of $M$
rewritten in the form
$\epsilon(\vec R,y)\equiv \epsilon(\vec r,x) M(\vec R,x,y)$.
Space averaging of this relation gives 
\begin{equation}
\overline{\epsilon}(1-\overline{M})=\overline{\left(\epsilon-
\overline{\epsilon}\right)\left(M-\overline{M}\right)}~,
\label{22}
\end{equation}
where $\overline{\epsilon}$ does not depend on $x$ (or $y$) 
due to the total flux conservation. 
If correlations between $\epsilon$ and $M$ were absent,
the right hand side of (\ref{22}) would be zero. It would
imply $\overline{M}=1$ and $\mu_{n,1}=0$, in agreement
with eq.(\ref{3}) that neglects considered correlations.
Positive correlations between $\epsilon$ and $M$,
i.e., a positive right-hand side of (\ref{22}), implies that 
$\overline{M}<1$, which means in turn $\mu_{n,1}<0$.

According to \cite{vm}, the $q$-dependence of the intermittency
exponents is given by
\begin{eqnarray}
&\mu_{n,q}=\gamma_n\, q \left[{\rm arsh}\left(q/a_n\right)-
{\rm arsh}\left(1/a_n\right)\right]+\mu_{n,1}\, q~,
\label{23}\\
&q\leq q_{n,M}={3/\gamma_n}\left[ 1/2 +\left(1/4+
  \left(a_n\gamma_n/3\right)^2 \right)^{1/2} \right]^{1/2}~\nonumber
\end{eqnarray}
(which corresponds to the dimensionality
$D_{n,q}\equiv 3+\mu_{n,q} -q \mu'_{n,q}=
3-\gamma_n\, q^2/\sqrt{q^2+a_n^2}$, so that $D_{n,q_{n,M}}=0$).
It can be verified now that these formulas (have not been
used above except the case of infinite $n$, eq.(\ref{15})) are
well-compatible with the current theory. This allows one to
extend the predicted $\mu_{n,q}$ for all positive $q$.
For $n=1$,  an extra constrain of the kind (\ref{18})
should hold:
\begin{equation}
F(\alpha_{1,q_{1,M}},\alpha_{1,0})\approx 0\quad 
(\alpha_{n,q}\equiv \mu'_{n,q})~.
\label{24}
\end{equation}
It allows one to evaluate the single unknown parameter $\mu_2$ in formula
(\ref{21}) for $\mu_{n,q}$. Specifically, minimization of the norm of
relative deviation between right-hand sides of (\ref{21}) and (\ref{23})
for $n=1$, subject to the constrain (\ref{24}), gives $\mu_2\approx 0.215$
for the intermittent turbulence.

\bigskip  
$\bullet$ {\bf COMPARISON WITH EXPERIMENT:}\break
The deviation of the above theoretical $\mu_{n,q}$ from the experimental
data of \cite{ppn} does not exceed $\delta\, \mu_q$, where 
$\delta\approx\mu_2^2/2\approx 0.023$ is the small parameter.
The deviation contains regular oscillations in $n$, which sweep is of
the order of  $\delta\, \mu_q$ (see Fig.1). 
This indicates that there is an additional
small effect that has not been taken into account above.  
It consists in some oscillations in the
probability of blow-up branching (i.e., hot spot splitting) as
the function of  $n$. Stronger branching makes larger the local 
dimensionality $D_{n,q}$ 
and reduces the local amplification $M$. The regions
of stronger branching repeat
with some period in $n$, which implies a correlation 
between consequent $M$. Calculation of
such a correlation should improve the accuracy of
theory (and probably put it beyond
the  accuracy of currently available experimental data).

\bigskip  
$\bullet$ {\bf SUMMARY:}\indent
This work has far reaching
theoretical implications. In particular, it indicates that the
intermittency problem can be solved by appropriate perturbation
methods. The reason  is that the major  correlations
causing intermittency can be described explicitly, while the
remaining correlations are small and can be treated as perturbations.
More specifically, it is shown that exactly solvable model (\ref{3}),(\ref{5})
can be appropriately ``dressed" (see eq.(\ref{21}) with the environment)
to describe very well the experimentally measured  
intermittency exponents $\mu_{n,q}$ at all $n$ and $q$ for which
these values are currently available.
Noteworthy, that the equations derived 
for scale-dependent intermittency exponents
$\mu_{n,q}$ do not contain
any adjusting parameter (that should be determined from comparison
with experiment), but do have both the Kolmogorov and the intermittent
solutions.

\bigskip  
$\bullet$ {\bf ACKNOWLEDGMENTS:}\indent
Research supported in part  by grants of Alfred P.~Sloan Foundation
and NEC Research Institute, Inc.. I am thankful to Profs.
G.~Pedrizzetti, E.A.~Novikov and A.A.~Praskovsky for sending 
me their data \cite{ppn}; to Prof. K.R.~Sreenivasan for consideration
of experimental methods, to Acad. B.~Chirikov,  
Profs. B.~Konopelchenko, A.~Milstein, T.~Spencer 
and Dr. Yu.~Tzidulko for useful discussions.

\end{document}